# Spectral Methods for Immunization of Large Networks

**Muhammad Ahmad**
14030004@lums.edu.pk
Department of Computer Science, School of Science and Engineering, Lahore University of Management Sciences

**Juvaria Tariq**
14070004@lums.edu.pk
Department of Computer Science, School of Science and Engineering, Lahore University of Management Sciences

**Mudassir Shabbir**
mudassir.shabbir@itu.edu.pk
Department of Computer Science, Information Technology University

**Imdadullah Khan**
imdad.khan@lums.edu.pk
Department of Computer Science, School of Science and Engineering, Lahore University of Management Sciences

## Abstract

Given a network of nodes, minimizing the spread of a contagion using a limited budget is a well-studied problem with applications in network security, viral marketing, social networks, and public health. In real graphs, virus may infect a node which in turn infects its neighbor nodes and this may trigger an epidemic in the whole graph. The goal thus is to select the best $k$ nodes (budget constraint) that are immunized (vaccinated, screened, filtered) so as the remaining graph is less prone to the epidemic. It is known that the problem is, in all practical models, computationally intractable even for moderate sized graphs. In this paper we employ ideas from spectral graph theory to define relevance and importance of nodes. Using novel graph theoretic techniques, we then design an efficient approximation algorithm to immunize the graph. Theoretical guarantees on the running time of our algorithm show that it is more efficient than any other known solution in the literature. We test the performance of our algorithm on several real world graphs. Experiments show that our algorithm scales well for large graphs and outperforms state of the art algorithms both in quality (containment of epidemic) and efficiency (runtime and space complexity).

# 1 Introduction

Consider a *large* network of hospitals distributed, geographically, over a region. Due to presence of an active pathogen, there is a danger of pandemic in the region. We assume that pathogen can easily travel between *linked* hospitals. We are concerned with keeping the maximum of the hospital network clean of pandemic. Based on their geographic vicinity (or some other criteria), it is known for every pair of hospitals $X, Y$ whether contamination at $X$ forebodes certain contamination at $Y$. If this is true about some pair of hospitals $X, Y$, we call $X$ and $Y$ as linked (i.e. $X$ and $Y$ is an edge in the network graph).

Now a team of scientists have developed a vaccine for the pathogen. Unfortunately, due to cost or production constraints the vaccine is only available in limited quantity and only a small fraction of hospitals can receive it. It is assumed that once a hospital receives the vaccine it cannot be contaminated nor can it contaminate any other hospital. Given the scarcity of the vaccine resource it is a very serious question to ask whether a given hospital should or should not receive the vaccine so as to minimize the overall spread of the pathogen in the region. This question, in essence, is the topic of this work, which appears in various other scenarios. An efficient solution to this problem can be applied to diverse array of high-impact applications in public health. It will also be an important tool in cyber-security solutions. Finding important players in social networks is a fundamental problem in viral marketing, online advertisement and social networks monitoring. In its essence, the problem of finding important nodes in a network can be reduced to the main problem of this paper.

This is known, in the literature, as the *Network Immunization Problem*. The problem is studied in terms of graphs where each node represents a hospital (or any other resource-hungry entity) and an edge between a pair of nodes represents the *link* between the pair of hospitals. We use the *Susceptible−Infected−Susceptible (SIS)* model of infection spread in the network that is detailed in section 3. As in (Chen 2016), we formulate the problem as follows.

**Problem 1.** *Given a simple undirected graph $G = (V, E)$ and an integer $k$ find a set $S$ of $k$ nodes such that if we "immunize" $S$, renders $G$ least "vulnerable" to an attack over all choices of $S$.*

Clearly, the problem is ill-defined, unless a precise and quantifiable definition of the *vulnerability* of a network is provided. To this end, it turns out that the largest eigenvalue of the adjacency matrix of the graph is a good measure of vulnerability of the graph (Chakrabarti 2008).

## 1.1 Problem Formulation and solution approach

For a simple undirected graph $G = (V, E)$ with $|V| = n$ and $|E| = m$, the adjacency matrix $A_G$ (or just $A$ whenever the graph $G$ is clear from the context) is an $n \times n$ binary matrix with columns and rows representing the vertices of $G$ and cell entries representing the edges i.e. $A_G(i, j) = 1$ if and only if there is an edge from vertex $i$ to vertex $j$. Eigen spectrum, $\{\lambda_i\}_{i=1}^{n}$, of adjacency and Laplacian matrices has been studied a lot for certain graph properties (Chung 1997).

Of particular interest to us is the largest eigenvalue (denoted by $\lambda_1(A)$ or $\lambda_1(G)$ or $\lambda_1$ when G is obvious from the context) of the adjacency matrix also referred to as the first eigenvalue in literature. Epidemic threshold is an intrinsic property of a network. It is well known in the epidemiology literature that if the "virus strength" is more than the network's epidemic threshold, then an outbreak will occur. The epidemic threshold of an undirected graph G is directly proportional to $\lambda_1(G)$ (Chakrabarti 2008). For this reason in the epidemiology community, the largest eigenvalue of graphs is used to study mathematical models of diseases spreads and outbreaks (Ganesh 2005). In general $\lambda_1$ is related to the global connectivity of a graph. For example we know that

$$\Delta \geq \lambda_1 \geq d_{avg}$$

where $\Delta$ and $d_{avg}$ are the maximum and average degrees in the graph, respectively (West 2001). In the context of the discussion above it seems only logical to delete/immunize the vertices in $G$ such that the remaining graph has minimum possible largest eigenvalue.

Let $G = (V, E)$ be a graph and let $A$ be it's adjacency matrix. For a subset of vertices $S \subseteq V$, we denote by $G_{-S}$ the subgraph induced on vertices in $V \setminus S$. Similarly, for $S \subseteq V$, we define $A_{-S}$ to be the submatrix of $A$ obtained by deleting rows and columns in $A$ at corresponding indices in $S$. By definition of induced subgraph, it is clear that $A_{-S}$ is the adjacency matrix of $G_{-S}$. We will denote by $G_S$ the subgraph of $G$ induced by vertices in S and $A_S$ denotes the adjacency matrix of $G_S$. The precise formulation of the above problem is given as follows:

***Problem 2.*** *Given a simple undirected graph $G = (V, E)$ find a set S of k vertices so that $\lambda_1(A_{-S})$ is the minimum possible over all choices of $S \subseteq V$.*

Since $\lambda_1(G)$ can be computed in $O(m)$ time (Chen 2016), the natural algorithm to find the optimal set $S$ takes time $O\left(\binom{n}{k} m\right)$, which is exponential in $k$. Indeed Lemma 1 asserts that the general the problem is intractable (see Section 2 for a proof).

***Lemma 1.*** *Problem 2 is NP-Hard.*

In this work we propose a novel algorithm to approximately solve Problem 2. Our algorithm is quite intuitive, promises a better approximation compared to known approaches, and is easy to implement. We propose a combinatorial formulation of the problem in which we construct a subset of vertices based on a defined shield value. Our shield value measures the contribution of the vertices towards the largest eigenvalue of the graph. Definition of our shield value uses basic theoretical tools from spectral graph theory and relies on simple combinatorial interpretation of largest eigenvalue. Removing vertices with high shield values would yield optimal immunization results. We provide an efficient approximation algorithm to estimate the shield values of vertices.

We provide detailed analysis of running time and space complexity of our algorithm. We also empirically evaluate our algorithm on several real world data sets. The results show that our algorithm achieves much better immunization performance compared to previously known solutions. Moreover, ideas developed in this work are general and can have many other potential applications. Indeed this method has been successfully used for estimating the *spectral radius* of large graphs (Abbas 2017).

Rest of the paper is organized as follows. In Section 2 we provide a detailed background to Problem 2 and discuss its computational intractability and approaches to approximate it. An extensive literature review of the immunization problem is provided in Section 3. We propose our algorithm in Section 4 along with its approximation guarantees and complexity analysis. Section 5 contains results of experimentation on real world graphs. This section also provides comparison of performance of our algorithm with other known algorithms. Afterwards we conclude with a discussion on future directions. A preliminary version of this paper has already been appeared in (Ahmad 2016).

## 2   Background

As argued earlier, the brute-force solution, that checks the eigendrop achieved by each possible subset of size $k$ and selects the subset achieving the largest eigendrop, has runtime $O\left(\binom{n}{k} m\right)$, where $m$ is the number of edges in the graph. This running time is exponential in size of input ($k$). We prove that the problem under consideration is NP-Hard.

***Proof of lemma 1:*** A simple reduction from Minimum Vertex Cover Problem goes as follows:

If there exist $k$ vertices that cover all the edges in the graph, then from the following fact, deleting those vertices will imply that $\lambda_1(A_{-S}) = 0$.

***Fact 1.*** *Empty graph (graph with no edges) on $n$ vertices has eigenvalue zero with multiplicity $n$.*

Conversely if there exists a $k$-set $S$ of vertices such that $\lambda_1(A_{-S}) = 0$, then $S$ must be a vertex cover; since otherwise it contradicts the following implication of the Perron-Frobenius theorem.

***Fact 2.*** *(Serre 2002) Deleting any edge from a simple connected graph $G$ strictly decreases the largest eigenvalue of the corresponding adjacency matrix.*

For a reduction from Max Independent Set problem that doesn't use Perron-Frobenius theorem see appendix of (Chen 2016). Although Problem 2 is NP-Hard, the following greedy algorithm guarantees a $\left(1 - \frac{1}{e}\right)$-approximation to the optimal solution to Problem 2.

Approximation guarantee of GREEDY-1 follows from Theorem 1.

---

**Algorithm 1 : GREEDY-1 $(G, k)$**

---

$S \leftarrow \emptyset$

while $|S| < k$

$\quad v \leftarrow \underset{v \in V \setminus S}{\mathrm{argmax}} \left(\lambda_1(A_{-S \cup \{v\}})\right)$

$\quad S \leftarrow S \cup \{v\}$

return $S$

---

***Theorem 1.*** *(Nemhauser 1978) Let $f$ be a non-negative, monotone and sub-modular function, $f: 2^\Omega \to \mathbb{R}$. Suppose $\mathbb{A}$ is an algorithm, that choose a $k$ elements set $S$ by adding an element $u$ at each step such that $u = \arg\max_{x \in \Omega \setminus S} f(S \cup \{x\})$. Then $\mathbb{A}$ is a $\left(1 - \frac{1}{e}\right)$-approximate algorithm.*

In (Chen 2016) each subset was assigned a score, called *shield-value* that was meant to approximate eigendrop achieved by that set. For a set $S$ of size $k$, shield value, $Sv(S)$ can be computed in $O(k^2)$ when the eigenvector corresponding to largest eigenvalue is known. Hence, the straightforward method of finding the set with largest shield value takes $O\left(\binom{n}{k} k^2\right)$ time. Using the fact that the objective function based on shield value is sub-modular, (Chen 2016) gave a greedy algorithm with runtime $O(nk^2 + m)$. By Theorem 1 their algorithm guarantees a $\left(1 - \frac{1}{e}\right)$-approximation to the optimal shield value.

## 3   Related Work

While the focus of this paper is to target node immunization problem using spectral graph theoretic techniques, there is vast amount of literature on this problem with approaches from diverse areas of the subject. Initially in 2003 Brieseneister, Lincoln and Porras (Briesemeister 2003) studied the propagation styles of viruses in communication networks. Along with this, the effects of graph topology in the spread of an epidemic are described in (Ganesh 2005) and the conditions under which an epidemic will eventually die out are discussed. Similarly Chakrabarti et. al in (Chakrabarti 2008) devise a nonlinear dynamical system (NLDS) to model virus propagation in communication networks. They use the idea of *birth rate* $\beta$, *death rate* $\delta$, and *epidemic threshold* $\tau$, for a virus attack where birth rate is the rate with which infection propagates, death rate is the node curing rate and epidemic threshold is a value such that if $\frac{\beta}{\delta} < \tau$, infection will die out quickly else if $\beta/\delta > \tau$ infection will survive and will result in an epidemic. Virus propagation is studied in both directed and undirected graphs. For undirected graphs, they prove that epidemic threshold $\tau$ equals $1/\lambda$, where $\lambda$ is the largest eigenvalue of the graph. Thus for a given undirected graph, if $\beta/\delta > 1/\lambda$, then the epidemic will die out eventually. But none of these specifically discuss the graph immunization problem.

Afterwards the problem is studied using the edge manipulation scheme. In (Kuhlman 2013) dynamical systems are used to delete appropriate edges to minimize contagion spread. While Tong et al. in (Tong 2012) remove $k$ edges from the graph in a manner that eigendrop (difference in largest eigenvalues of original and resultant graphs) is maximized. For this edges are selected on the basis of left and right eigenvectors of leading eigenvalue of the graph such that for each edge $e_x$, $score(e_x)$ is the dot product of the left and right eigenvectors of leading eigenvalue of the adjacency matrix of $A$.

Graph vulnerability is defined as measure of how much a graph is likely to be affected by a virus attack. As in (Tong 2012), the largest eigenvalue is selected as a measure of graph vulnerability, in (Chen 2016) they also use largest eigenvalue for the purpose but instead of removing edges, nodes are deleted to maximize the eigendrop. Undirected, unweighted graphs are considered and nodes are selected by an approximation scheme using the eigenvector corresponding to largest eigenvalue which cause the maximum drop.

Zhang et al. and Song et al. adapt the non-preemptive strategy i.e. selection of nodes for immunization is done after the virus starts propagating across the graph. For this they use discrete time model to obtain additional information of infected and healthy nodes at each time stamp. In (Song 2015) directed and weighted graphs are used in which weights represent the probability of a healthy node being infected by its neighbors and node selection is done on the basis of these probabilities. Then results are evaluated on the basis of save ratio (SR) which is the ratio between the number of infected nodes when $k$ nodes are immunized over the number of infected nodes when no node is immunized. (Zhang 2014a) and (Zhang 2014b) consider undirected graphs and incorporates dominator trees for selecting nodes. Results are evaluated in terms of expected number of remaining infected nodes in the graph after the process of immunization.

In filter placement (Erdös 2012), those nodes are identified which are cause of maximum information multiplicity. Moreover some reverse engineering techniques are also used for similar problems. Prakash et al. (Prakash 2012) study the graphs in which virus has already spread for some time and they point out those nodes from where the spread started. From this they find out the likelihood of other nodes being affected.

Another direction to look at the problem is to consider graphs in which some nodes are already infected and these nodes can spread virus among other reachable nodes or graphs in which all nodes are contaminated and the goal is to decontaminate the graph by using some agent nodes which traverse along the edges of the graph and clean the nodes. The problem is usually referred to as decontamination of graph or graph searching problem. Different models are studied to solve the problem and most of them assume the monotonicity in decontamination i.e. once a node is decontaminated then it cannot get contaminated again (Bienstock 1991), (Flocchini 2008), (Flocchini 2007), (Fraigniaud 2008). But non-monotonic strategies are also studied (Daadaa 2016).

Other work that is related to node immunization is the selection of most influential nodes in a given network to maximize the diffusion of new information in a network. Kempe et al. provided the provably efficient approximation algorithm for the problem (Kempe 2003). Seeman and Singer (Seeman 2013) use stochastic optimization models to maximize the information diffusion in social networks. Influence maximization problem is slightly different from immunization problem as in influence maximization problem the goal is to select nodes for seeding which will maximize the spread on new idea while in node immunization problem goal is to select nodes which will help in minimal spread of virus.

# 4 Our Proposed Algorithm

In (Chen 2016) they defined shield value that was an approximation to the eigendrop. We define our shield value to be the eigendrop and approximate the eigenvalue computation instead. We achieve better theoretical guarantees with our shield value definition. Furthermore, it is easy and computationally efficient to approximate our shield value.

## 4.1 Our shield value and its justification

We use the following well known facts from linear algebra and graph theory. Given an $n \times n$ matrix A, let trace of $A$ be denoted by $tr(A)$,

*Fact 3.* [c.f. (Strang 1976)]

$$tr(A) = \sum_{i=1}^{n} a_{ii} = \sum_{i=1}^{n} \lambda_i(A)$$

*Fact 4.* [c.f. (West 2001) p. 455 ]

$$tr(A^p) = \sum_{i=1}^{n} \lambda_i(A^p) = \sum_{i=1}^{n} \lambda_i(A)^p$$

Clearly, for even powers $p$, $tr(A^p)$ is an upper bound on $\lambda_1^p$ and as $p$ grows $tr(A^p)$ approaches $\lambda_{max}^p(A)$ where $\lambda_{max} = \max\{|\lambda_i(A)|\}$. We therefore, try to remove a set $S$ of $k$ vertices from the graph such that $tr((A_{-S})^p)$ is minimized. The goodness of a set $S$ in this setting is defined as

$$f_p(S) = tr((A_{-S})^p) \tag{1}$$

Define

$$g_p(S) = tr(A^p) - tr((A_{-S})^p) \tag{2}$$

Note that minimizing $f_p(S)$ is same as maximizing $g_p(S)$.

Given a graph $G = (V, E)$, a walk $W$ of length $l$ in $G$ is a sequence of vertex $v_0, v_1, \ldots, v_l$ such that for $0 \leq i \leq l-1$ $(v_i, v_{i+1}) \in E$. We say that $W$ is a walk from $v_0$ to $v_l$. If $v_0 = v_l$, then $W$ is called a closed walk. For $v \in V$, let $\mathbb{CW}_i(G, v)$ be the set of closed walks of length $i$ in $G$ containing the vertex $v$. Suppose $CW_i(G, v) = |\mathbb{CW}_i(G, v)|$. When $X \subseteq V$, $\mathbb{CW}_i(G, X) = \bigcup_{x \in X} \mathbb{CW}_i(G, x)$. We similarly denote by $CW_i(G, X)$ to be the cardinality of the set $\mathbb{CW}_i(G, X)$. When $X = V$, we refer to $\mathbb{CW}_i(G, X)$ as $\mathbb{CW}_i(G)$. So $\mathbb{CW}_i(G)$ is the set of all closed walks of length $i$ in $G$.

We use the following well-known fact from graph theory

**Fact 5.** [c.f. (West 2001) p.455] *Given a graph $G$ with adjacency matrix $A$,*

$$CW_p(G) = tr(A^p) = \sum_{i=1}^{n} \lambda_i^p.$$

It is clear from the definitions and fact 5 that for $\subset V$,

$$CW_p(G) = CW_p(G_{-S}, V \setminus S) + CW_p(G, S). \tag{3}$$

This identity is same as the one already mentioned in (3). This similarly suggests a strategy namely one should delete a set $S$ of $k$ vertices such that $CW_p(G_{-S}, V \setminus S)$ is minimized for a large even integer $p$ (equivalently $CW_p(G, S)$ is maximized). The goodness of a set of vertices $S \subseteq V$, in terms of number of closed walks is given by

$$g_p(S) = CW_p(G, S) \tag{4}$$

which as noted above we would like to maximize. Hence we reduced the problem of finding a set maximizing the eigendrop to the problem of finding a set of vertices that are part of many closed walks. We note that this latter problem is of a combinatorial nature and more amenable to techniques from graph theory.

**Algorithm 2 : GREEDY-2 ($G, k, p$)**

```
S ← ∅
while |S| < k do
    v ← argmax CW_p(G_S, v)
         v∈V\S
    S ← S ∪ {v}
return S
```

An algorithm based on this intuition is given in Algorithm GREEDY-2$(G, k, p)$. Quality guarantee of Algorithm GREEDY-2$(G, k, p)$ follows from sub-modularity of the optimization function (4) and Theorem 1. Next we show that the objective function given in (4) is non-negative, monotone sub-modular, i.e. we show that $g_p(S)$ has the property of diminishing return.

**Lemma 2.** *The function $g_p(S)$ given in (4) is non-negative, monotonically non-decreasing and sub-modular.*

*Proof.* Since $g_p(S)$ counts the number of closed walks of length $p$, it is clearly non-negative. By definition of $g_p(S)$ for $X \subseteq Y$ we have

$$g_p(Y) - g_p(X) = CW_p(G, Y) - CW_p(G, Y)$$

$$= CW_p(G, (Y \setminus X) \cup X) - CW_p(G, X)$$

$$= CW_p(G, X) + CW_p(G_{-X}, Y \setminus X) - CW_p(G, X)$$

$$= CW_p(G_{-X}, Y \setminus X) \geq 0$$

where the last inequality follows from non-negativity of $g_p(S)$. Hence $g_p(S)$ is monotonically non-decreasing function.

For sub-modularity of $g_p(S)$, let $X, Y, Z \subset V$ such that $X \subset Y$ and $Z \cap Y = \emptyset$. Let $L = X \cup Z$ and $R = Y \cup Z$. We have

$$\left(g_p(R) - g_p(Y)\right) - \left(g_p(L) - g_p(X)\right)$$

$$= CW_p(G, R) - CW_p(G, Y) - CW_p(G, L) + CW_p(G, X)$$

$$= CW_p(G, Y \cup Z) - CW_p(G, Y) - CW_p(G, X \cup Z) + CW_p(G, X)$$

$$= CW_p(G, Y) + CW_p(G_{-Y}, Z) - CW_p(G, Y) - CW_p(G, X) - CW_p(G_{-X}, Z) + CW_p(G, X)$$

$$= CW_p(G_{-Y}, Z) - CW_p(G_X, Z) \leq 0$$

where the last inequality follows from the fact that by definition $\mathbb{CW}_p(G_{-Y}, Z) \subseteq \mathbb{CW}_p(G_{-X}, Z)$, hence $CW_p(G_{-Y}, Z) \leq CW_p(G_{-X}, Z)$. □

For a vertex $v$, $CW_p(G, v)$ can be computed by the powers of $A$ i.e $A^2, A^3, \cdots, A^p$. Clearly computing all these matrices for large $p$ when $G$ is significantly large as well is computationally expensive. So the Algorithm GREEDY-2$(G, k, p)$ is practically impossible to be executed. So we set $p = 4$ to approximate the result and our practical experimentation shows that $p = 4$ gives good enough quality guarantee. Although we have also done approximation using $p = 6$ in (Tariq 2017), which certainly gives better results, we focus on $p = 4$ only in this paper.

First we give a closed form expression for $CW_4(G, v)$ in terms of degrees and codegrees. For a given graph $G$, $N_G(x) = \{y \in V(G): A_G(x, y) = 1\}$ and $d_G(x) = |N_G(x)|$. Define $N_G(x, y) = \{z \in V(G): A_G(x, z) = 1 \land A_G(y, z) = 1\}$ to be the common neighborhood of $x$ and $y$ in G. Let $d_G(x, y) = |N_G(x, y)|$, note that $N_G(x, x) = N_G(x)$ and $d_G(x, x) = d_G(x)$. When $G$ is clear in the context, we refer to $N_G(x, y)$ as $N(x, y)$ and similarly to $d_G(x, y)$ as $d(x, y)$.

***Lemma 3.*** *For any vertex $v \in V$,*

$$CW_4(G, v) = 2d(v)^2 + 4 \sum_{\{u \in V,\ u \neq v\}} d(u, v)^2.$$

*Proof.* A closed walk $W: (v, x, u, y, v)$ of length 4 can be interpreted as the concatenation of two walks of length 2 with same end points $u$ and $v$. The number of walks of length 2 with the end points $u$, $v$ is $A_G^2(u, v)$. We want to count the closed walks of length 4 that contain a fixed vertex $v$ atleast once. Note that $v$ can occur at most twice in a closed walk of length 4. In a closed walk of length 4 there are 4 positions for vertices (Since first and last vertex is same, we consider it one position).

First we count the closed walks of length 4 that contain $v$ exactly in one position. Call the set of closed walks of length 4 with v at $i_{th}$ position as $C_4(v, i)$. This gives the total number of closed walks of length 4 with appearance of v exactly once as $\sum_{\{i=1\}}^{4} |C_4(v, i)|$. Any closed walk in $C_4(v, 1)$ is of the form $(v, a, b, c, v)$, where $a, b, c$ are vertices in $V(G)$. Number of these walks is $\sum_{\{b \neq v\}} (A^2(v, b))^2$. Clearly $(c, v, a, b, c)$ represents any closed walk in $C_4(v, 2)$. Note that for a fixed $a, b$ and $c$, $(c, v, a, b, c)$ is one position clockwise rotation of closed walk $(v, a, b, c, v)$. This implies that corresponding to every closed walk in $C_4(v, 1)$ there is a closed walk in $C_4(v, 2)$ and vice versa. Hence the number of walks in $C_4(v, 2)$ is also $\sum_{\{b \neq v\}} (A^2(v, b))^2$. Similarly we get that the number of walks in $C_4(v, 3)$ and $C_4(v, 4)$ is also the same. So we have $\sum_{\{i=1\}}^{4} |C_4(v, i)| = 4 \sum_{\{u \neq v\}} (A^2(u, v))^2$.

Second we consider closed walks in which $v$ appears twice. Now it is possible in two ways: 1) $v$ appears in $1^{st}$ and $3^{rd}$ position as $(v, a, v, c, v)$ or 2) $v$ takes $2^{nd}$ and $4^{th}$ position as $(a, v, c, v, a)$. Clearly there are $2(A^2(v, v))^2$ such walks.

This gives total number of closed walks of length 4 containing a vertex $v$ as

$$CW_4(G, v) = 2(A^2(v, v))^2 + 4 \sum_{\{u \neq v\}} A^2(u, v)^2$$

which is same as what we required. □

We incorporate the above formula in the following algorithm. For a given vertex $v$, $score_G(v) = 4\sum_{\{u \in V(G)\}} d(u,v)^2 - 2d(v)^2$ then $\sum_{\{u \neq v\}} d(u,v)^2$ can be computed by taking the characteristic vector $\chi_v$ of $N(v)$ (a binary vector of length $n$ where the $\chi_v[i] = 1 \leftrightarrow (v, v_i) \in E$). Then for each $vertex\ u \in V \setminus \{v\}$ we go through each neighbor $x$ of $u$ in its adjacency list and check if $\chi_v[x] = 1$ to increment $d(v, u)$.

It takes $O(m)$ time to compute the $score_G(x)$ for a vertex $x$, to get the vertex with maximum score it takes $O(nm)$ time. Note that after removing $k$ vertices (for constant $k$) the graph still has $O(m)$ edges. Now we give an efficient approximation to $score_G(x)$ that not only can be computed in linear time but also can be updated after removing a vertex $y$ in time proportional to $d(x)$.

We have

$$\left(\frac{\sum_{\{u \neq v\}} d(u,v)}{n}\right)^2 \leq \frac{\left(\sum_{\{u \neq v\}} d(u,v)^2\right)}{n} \leq \left(\frac{\sum_{\{u \neq v\}} d(u,v)^2}{n}\right) \tag{5}$$

The first inequality is the Cauchy-Schwarz inequality, [c.f (Kreyszig 1989), while the second follows from the fact that $d(u,v)$ is non-negative for each $u, v$.

In view of the above inequality, we approximate the $score\_G(v)$ by $score\_G'(v)$ given as

$$score'_G(v) = 2d_G^2(v) + 4\left(\sum_{\{u \neq v\}} d_{G(u,v)}\right)^2. \tag{6}$$

Our motivation to use $score'_G(x)$ is that not only it is easy to compute but also after a vertex is deleted it is easy to update the scores of all vertices in the remaining subgraph.

## 4.2 Algorithm

Now we give algorithm to compute the results. First we show procedure to find $score'_G(v)$ for graph $G$ in Algorithm COMPUTE-SCORE($G$), then in Algorithm UPDATE-SCORE($G, v_i$) we give algorithm to update the scores of vertices when a certain vertex $v_i$ is deleted from the graph and finally we discuss the greedy approach to approximate that which vertices should be deleted in order to immunize the graph in Algorithm GREEDY-3($G, k$) and along with these we discuss the time and space complexity in this section. Now we give algorithm to compute $score'_G(v)$;

---

**Algorithm 3 : COMPUTE-SCORE (G)**

1: $deg \leftarrow ZEROS(n)$ ▷ *Initialize the degree array to n zeros*
2: $codegSum \leftarrow ZEROS(n)$ ▷ *Initialize the codegree sum array to n zeros*
3: $score'_G \leftarrow ZEROS(n)$ ▷ *Initialize all scores to zeros*
4: **for** each vertex $v_i$ **do**
5:     **for** each neighbor $v_j$ of $v_i$ **do**
6:         $deg_G[v_i] \leftarrow deg_G[v_i] + 1$
7: **for** each vertex $v_i$ **do**
8:     **for** each neighbor $v_j$ of $v_i$ **do**
9:         $codegSum[v_j] \leftarrow codegSum[v_j] + deg_G[v_i] - 1$
10: **for** each vertex $v_i$ **do**
11:     $score'_G[v_i] \leftarrow 2 * deg_G[v_i]^2 + 4 * codegSum[v_i]^2$

---

**Lemma 4.** *Runtime of Algorithm COMPUTE-SCORE(G) is $O(m)$.*

*Proof.* It is clear that line 6 of Algorithm COMPUTE-SCORE($G$) takes $O(1)$ time and it is executed $O(m)$ times. Since loop at line 5 is iterated over all neighbors of a fixed vertex, $v_i$. Hence for $v_i$, line 6 is executed $d_G(v_i)$ times. Thus for all $v_i \in G$, line 6 runs for $\sum_{\{v_i \in V(G)\}} d\_G(v\_i) = 2m$ (Bondy 1976). Same is true for line 9.

Line 11 has constant time computation while it is computed for every vertex, thus it takes $O(n)$ time. So total time taken to compute score of every vertex is $O(m)$. □

For a given vertex $v$ of $G$, we update the score of vertices after removing $v$ in the following way;

**Algorithm 4 : UPDATE-SCORE ($G,v_i$)**

1: **for** each neighbor $v_j$ of $v_i$ **do**
2:   $\deg_G[v_j] \leftarrow \deg_G[v_j] - 1$
3:   $codegSum[v_j] \leftarrow codegSum[v_j] - (\deg_G[v_i] - 1)$
4:   **for** each neighbor $v_k$ of $v_j$ **do**
5:     $codegSum[v_k] \leftarrow codegSum[v_k] - 1$
6: $\deg_G[v_i] \leftarrow 0$
7: $codegSum[v_i] \leftarrow 0$
8: $score'_G[v_i] \leftarrow 0$
9: **for** each neighbor $v_j$ of $v_i$ **do**
10:   $score'_G[v_j] \leftarrow 2 * \deg_G[v_j]^2 + 4 * codegSum[v_j]^2$
11:   **for** each neighbor $v_k$ of $v_j$ **do**
12:     $score'_G[v_k] \leftarrow 2 * \deg_G[v_k]^2 + 4 * codegSum[v_k]^2$

**Lemma 5.** *Algorithm UPDATE-SCORE($G, v_i$) takes $O(m)$ time to update score with respect to parameter $v_i$.*

*Proof.* In algorithm, line 2, 3 and 8 takes constant time steps and both are computed $d_G(v\_i)$ times. However line 5 and 12 are computed $\sum_{\{v_j \in N_G(v_i)\}} d_G(v_j)$ times which is upper bounded by $m$. Hence for a fixed vertex $v_i$ runtime of algorithm is $O(m)$. □

We here give the proof of correctness of the Algorithms COMPUTE-SCORE($G$) and UPDATE-SCORE($G, v_i$).

**Lemma 6.** *For each vertex $v \in V(G)$ Algorithm COMPUTE-SCORE($G$) computes the $score'_G(v)$ as defined in (5).*

*Proof.* It is clear that the Algorithm COMPUTE-SCORE($G$) computes the first term correctly, as each neighbor $v_i$ of v contributes 1 to the degree of $v$. To see why the second term is computed correctly, consider the following fact:

**Fact 6.** *For $\in V(G)$, $\sum_{\{u \neq v\}} d_G(u, v) = \sum_{\{w \in N_G(v)\}} (d_{G(w)} - 1)$*

*Proof.* The left hand side is counting all occurrences of all vertices w such that w is a common neighbor of $u$ and some vertex $v$. Essentially counting all paths of length 2 from $u$ to $v$ where $w$ is the center vertex.

We count these structures by counting the number of times each center vertex thus appear. Since $v$ is fixed, the number of times a vertex $w$ appears in such a structure is exactly the number of neighbors of $w$, i.e. $d_G(w)$ times. Now since $v$ is fixed and it is also a neighbor of $w$, we subtract one from it. Hence the expression on the right hand side follows. □

**Lemma 7.** *For all vertices $u, v \in V(G)$ Algorithm UPDATE-SCORE(G, $v_i$) computes the $score'_{G_{-\{u\}}}(v)$ as defined in (5).*

*Proof.* To see this, consider a vertex $v$ which is neighbor of $u$. We note that first contribution of $u$ in score' of $v$ is in the first term i.e. degree of $v$, so we decrease the degree of $v$ by one. Clearly from the figure below, $u$ adds $d_G(u) - 1$ in codegree sum of $v$, which is the second term of the $score'_G(v)$, given as $\sum_{\{u \neq v\}} d_G(u, v)$.

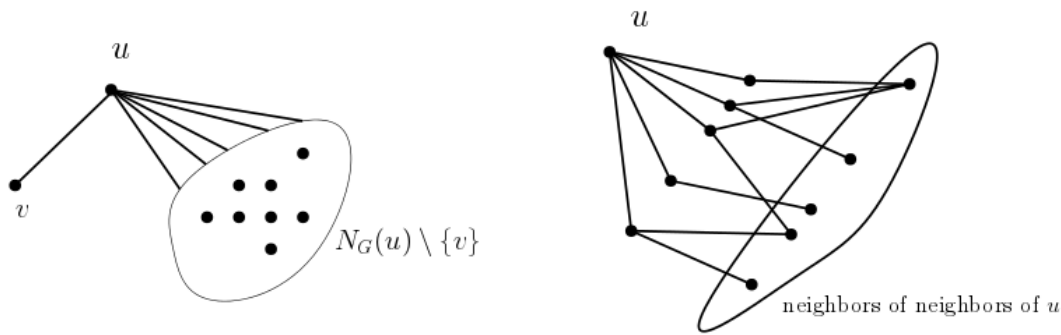

Now consider the vertices $v_i$, which are neighbors of neighbors of $u$. Score of these vertices is also effected by removing $u$ since $u$ contribute one to codegree sum of $v_i$. □

Here is the algorithm to select $k$ vertices in the given graph $G$ such that approximated eigendrop is maximized after deleting those vertices.

**Algorithm 5 : GREEDY-3 ($G, k$)**

1: $S \leftarrow \emptyset$
2: COMPUTE-SCORE($G$)
3: **while** $|S| < k$ **do**
4:     $v \leftarrow \underset{u \in V \setminus S}{argmax}\ score'_G[u]$
5:     $S \leftarrow S \cup \{v\}$
6:     UPDATE-SCORE($G,v$)
7: **return** $S$

**Theorem 2.** *The computational complexity of the Algorithm GREEDY-3($G, k$) is $O(n + km + k \log(n))$, while the space complexity is $O(m + n + k)$.*

*Proof.* By Lemma 4, line 2 takes $O(m)$ time. The scores for all vertices can be stored in a MAX-HEAP, which can be built in $O(n)$ time while each UPDATE-KEY and EXTRACT-MAX takes $O(\log(n))$ time (Cormen 2009). We extract max from the heap $k$ times, hence its total runtime is $O(\log(n))$. As argued by Lemma 5, time consumed in each call to UPDATE-SCORE takes $O(m)$ time, we get that total time taken by the algorithm is $O(n + km + k \log n)$. For the space complexity, in addition to storing the graph that takes $O(n + m)$ space, we need $O(n)$ space to store the three additional arrays.

## 5 Experimental Section

We present results of our approximation algorithm in this section. We evaluate the goodness of our algorithm on real world graphs to quantify the scalability and effectiveness on large graphs. Throughout this paper we have used *Susceptible−Infected−Susceptible (SIS)* model for virus propagation in graphs. For benchmark comparison we also implemented MAX-DEGREE and UPDATED-MAX-DEGREE. MAX-DEGREE picks the top $k$ maximum degree vertices for immunization while UPDATED-MAX-DEGREE selects the vertex with the maximum degree and deletes that vertex and repeats $k$ times. We use the NET-SHIELD implementation which is available online[1]. We implemented our proposed algorithm in Matlab and our implementation along with source code and documentation is available online on the given link [2].

| Name | Node Count | Edge Count |
|------|------------|------------|
| Karate | 34 | 78 |
| Oregon | 10,670 | 22,002 |
| AA | 418,236 | 2,753,798 |

Table 1 Summary of Datasets

The data sets used in experimentation are described in Table 1. All the real graphs used for experimentation obey power law distribution of degrees. The first data set is of a local karate club and is named as Karate graph[3]. Nodes of the graph represent members of the club and an edge between nodes show that corresponding members are friends with each other. Graph consists of 34 nodes and 78 edges. The graph is undirected and unweighted.

The second data set is from Oregon AS (Autonomous System)[4] router graphs, which are AS level connectivity networks inferred from Oregon route views. There are a number of Oregon AS graphs available and each node represents a router and an edge between two routers represents a direct peering relationship between two routers. We have selected one set from Oregon router graphs having 10,670 nodes and 22,002 edges. The graph is undirected and unweighted. Nodes selected by greedy algorithm are those routers whose immunization will maximally reduce the spread of virus.

The third data set (AA) is from DBLP[5] dataset. In graph a node represents an author and presence of an edge between two nodes shows that two authors have a co-authorship. In DBLP there is total node count of 418,236 and the number of edges among nodes is 2,753,798. We extracted smaller graphs by selecting co-authorship graph of only one journal (e.g Displays, International Journal of Computational Intelligence and Applications, International Journal of Internet and Enterprise Management, etc.). We ran our experiments on 20 different smaller co-authorship networks based on co-authorship graphs of 20 different journals. For the smaller sub graphs that we have extracted from DBLP dataset, node count goes up to few thousands and edge count goes up to few ten thousands. Co-authorship graph of ActaInf contains 1,791 nodes and 1,659 edges, graph of AI Communication journal has node count of 1,203 and edge count of 2,204 , sub graph of Asia-Pacific Journal of Operational Research (APJOR) has 1,132 nodes and 1,145 edges, Computer in Industry journal contains 2,844 nodes and 4,466 edges among nodes, journal of IEEE Wireless has total of 7,882 authors and 16,557 co-authorship links among authors. Detail of sub graphs of DBLP data set is also given in Table 2. These sub graphs are also undirected and unweighted.

---

[1] https://www.dropbox.com/s/aaq5ly4mcxhijmg/Netshieldplus.tar.
[2] www.dropbox.com
[3] http://konect.uni-koblenz.de/networks/ucidata-zachary
[4] http://snap.stanford.edu/data/oregon1.html
[5] http://dblp.uni-trier.de/xml/

| Name | Node Count | Edge Count |
|---|---|---|
| ActaInf | 1,791 | 1,659 |
| AI Communication | 1,203 | 2,204 |
| APJOR | 1,132 | 1,145 |
| Computer In Industry | 2,844 | 4,466 |
| Computing And Informatics | 1,598 | 2,324 |
| Ecological Informatics | 1,990 | 4,913 |
| IEEE Wireless | 7,882 | 16,577 |
| IJCIA | 848 | 975 |
| IJIEM | 373 | 357 |

Table 2 Summary of DBLP subgraphs

We did extensive experimentation with varying count of nodes to be immunized. In the results shown, x-axis shows the value of *k* which is count of immunized nodes and y-axis shows the percentage of eigendrop which is the achieved benefit after deleting *k* nodes from graph.

Results are evaluated on the basis of percentage of eigendrop. Eigendrop is difference of largest eigenvalues of original graph and perturbed version of graph after immunization of *k* nodes.

$$\Delta\lambda = \lambda - \lambda(S) \qquad (7)$$

where *S* is the set containing nodes to be immunized having cardinality *k*. It is clear from the results that our greedy algorithm outperforms NetSheild approach and other approaches like top *k* degree and updated maximum degree approach. Our greedy algorithm is scalable for larger values of *k* as well as for larger graphs as our algorithm has less running time complexity than NetSheild, top *k* degree and updated maximum degree.

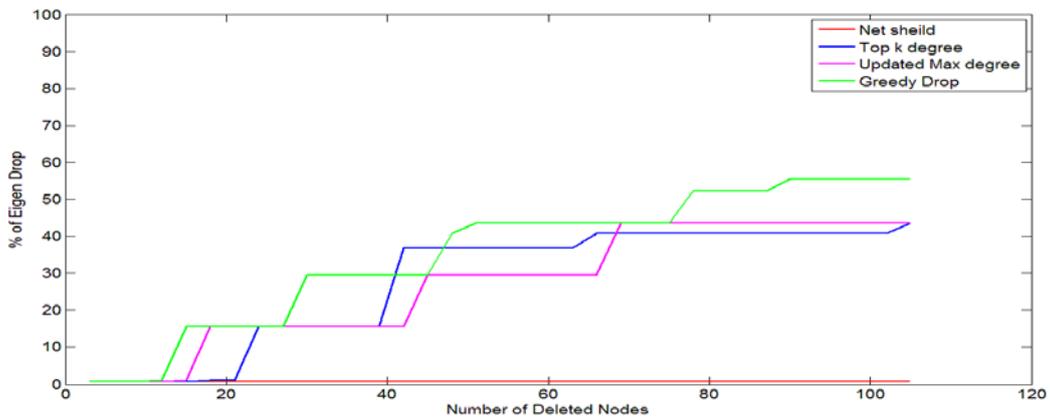

Figure 1 Eigendrop of Actainf Graph

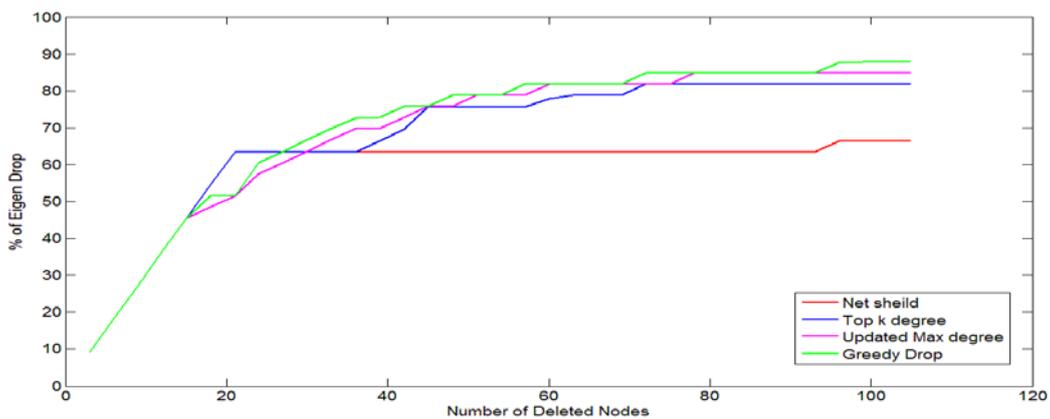

Figure 2 Eigendrop of AICommunication Graph

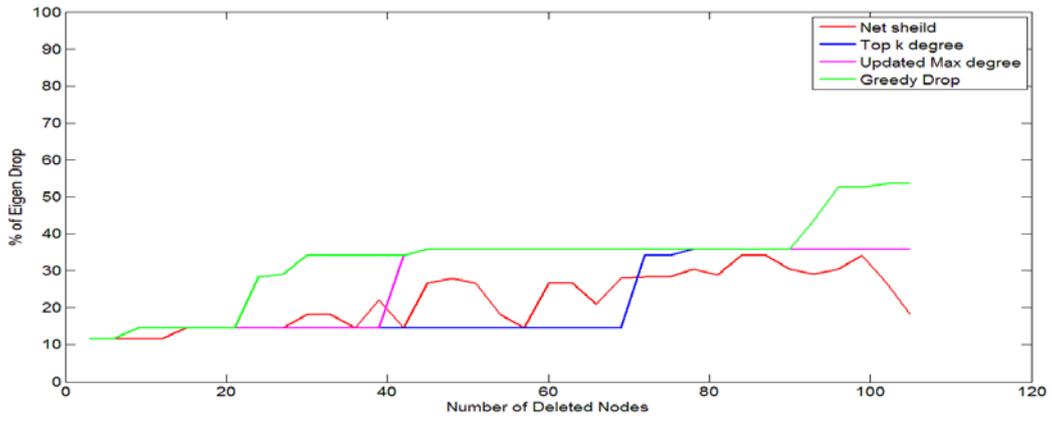

Figure 3 Eigendrop of APJOR Graph

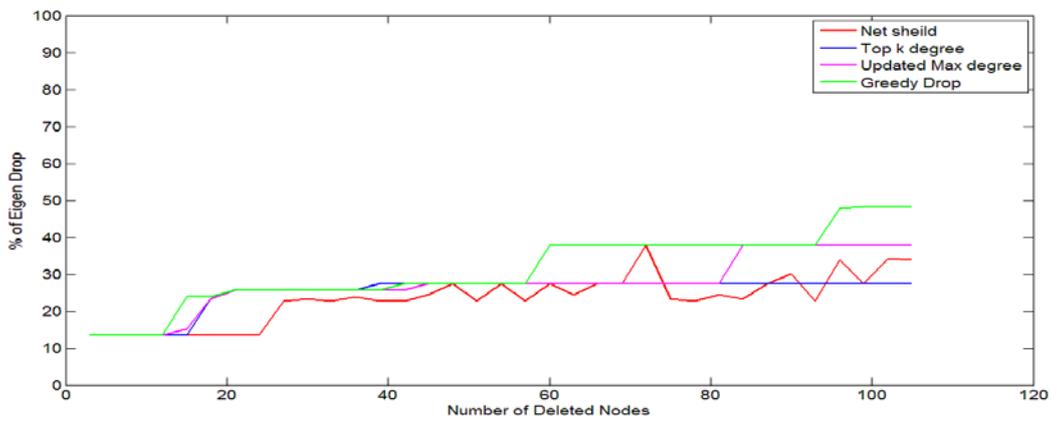

Figure 4 Eigendrop of Computer In Industry Graph

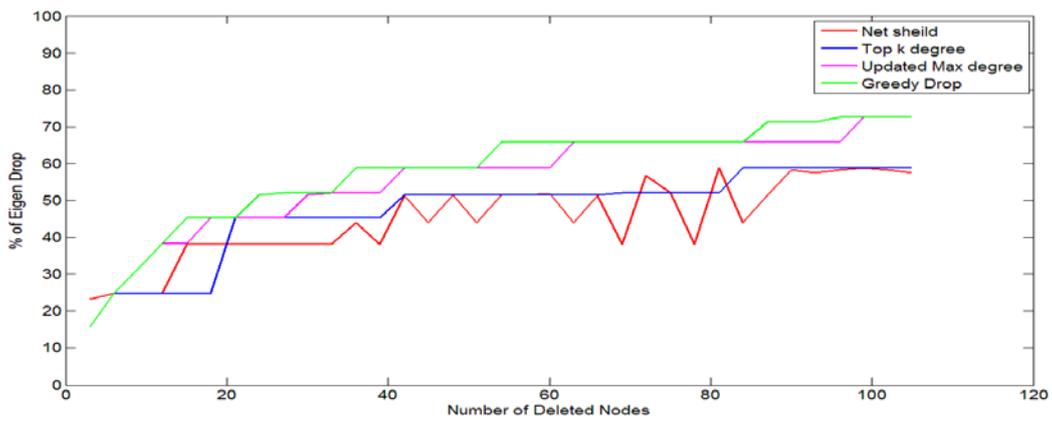

Figure 5 Eigendrop of Computing And Informatics Graph

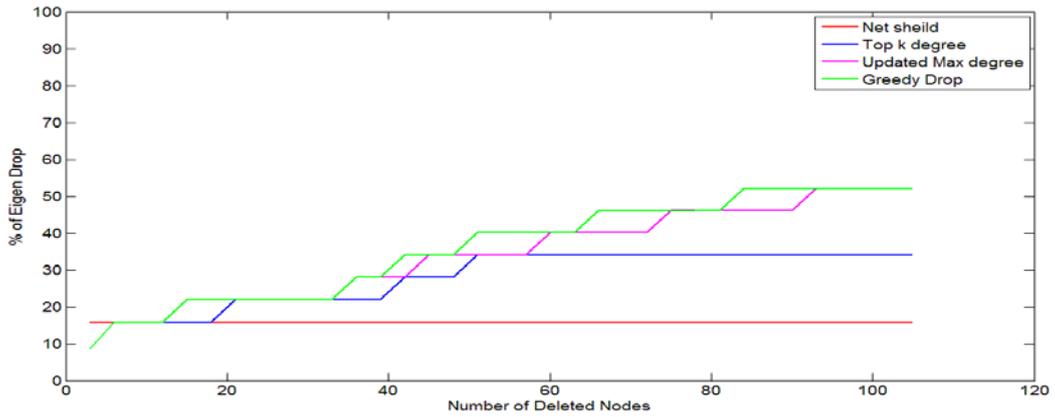

**Figure 6 Eigendrop of Ecological Informatics Graph**

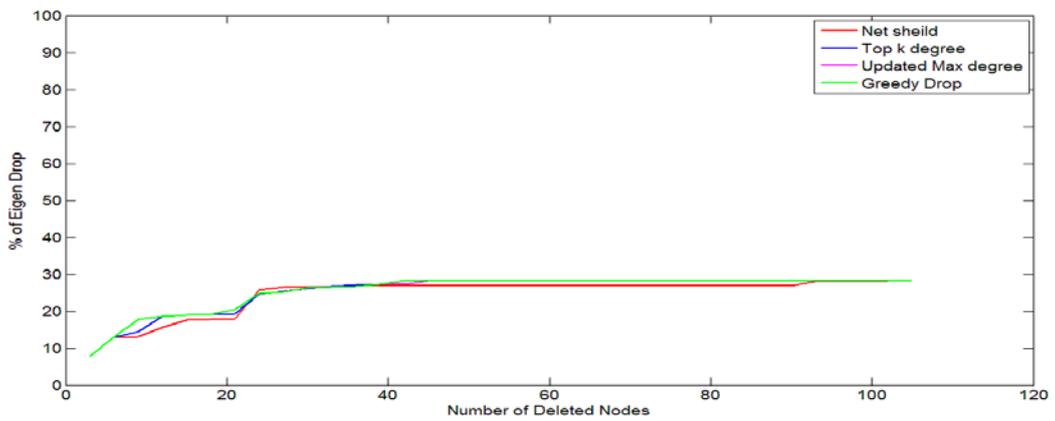

**Figure 7 Eigendrop of IEEE Wireless Graph**

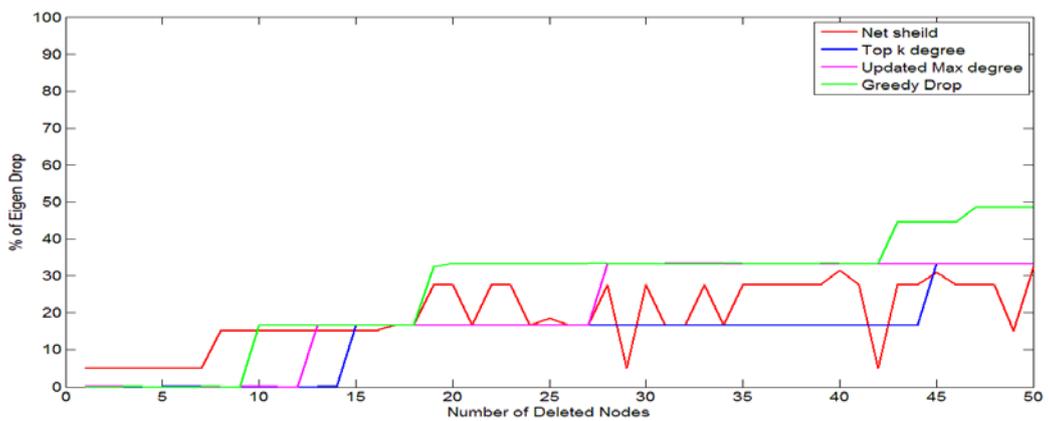

**Figure 8 Eigendrop of IJCIA Graph**

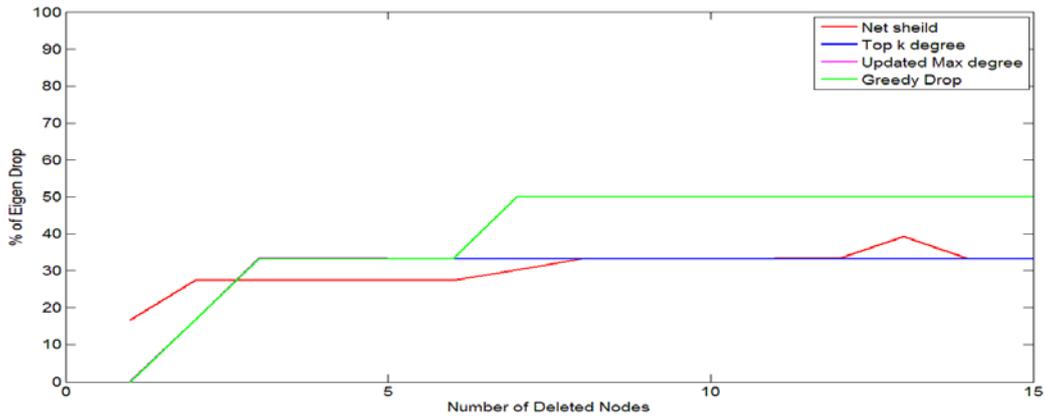

Figure 9 Eigendrop of IJIEM Graph

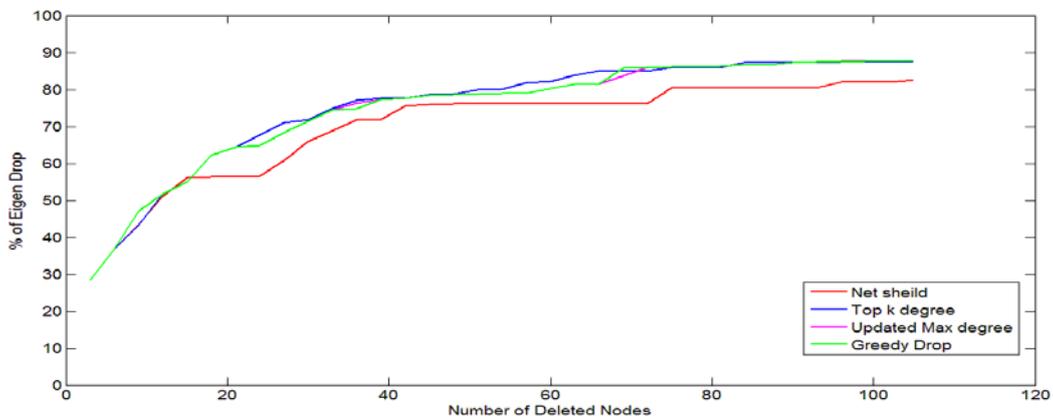

Figure 10 Eigendrop of Oregon Graph

# 6   Conclusion

In this work, we explored some links between established graph vulnerability measure and other spectral properties of even powers of adjacency matrix of the graph. We define shield value in terms of trace of the adjacency matrix of the graph. Based on these insights we present a greedy algorithm that iteratively selects $k$ nodes such that the impact of each node is maximum in the graph, in the respective iteration, and thus we maximally reduce the spread of a potential infection in the graph by removing those vertices. Our algorithm is scalable to large graphs since it has linear running time in the size of the graph. We have conducted experiments on different real world communication graphs to confirm the accuracy and efficiency of our algorithm. Our algorithm outperforms the state of the art algorithm in performance as well as in quality.

The main limitation of this work is that we used only $p = 4$, but it is theoretically clear that using large values of $p$ would yield better immunization performance. For the future work, we will consider the large values of the parameter $p$, used in the definition of our shield value.